# Expansion of situations theory for exploring shared awareness in human-intelligent autonomous systems

## Scott A. Humr* and Mustafa Canan

Information Science Department,
Naval Postgraduate School,
Monterey, CA 93943, USA
Email: scott.humr@nps.edu
Email: mustafa.canan@nps.edu
*Corresponding author

## Mustafa Demir

Ira A. Fulton Schools of Engineering,
Arizona State University,
Mesa, AZ 85212, USA
Email: mdemir@asu.edu

**Abstract:** Intelligent autonomous systems (IAS) are part of a system of systems (SoS) that interact with other agents to accomplish tasks in complex environments. However, IAS-integrated SoS add additional layers of complexity based on their limited cognitive processes, specifically shared situation awareness (SSA) that allows a team to respond to novel tasks. IAS's lack of SSA adversely influences team effectiveness in complex task environments, such as military command-and-control. A complmentary approach of SSA, called situations theory, is beneficial for understanding the relationship between SoS's SSA and effectiveness. The current study elucidates a conceptual discussion on situations theory to investigate the development of an SoS's shared situational awareness when humans team with IAS agents. To ground the discussion, the reviewed studies expanded situations theory within the context of SoS that result in three major conjectures that can be beneficial to the design and development of future SoSs.

**Keywords:** artificial intelligence; human-machine interaction; IAS; intelligent autonomous systems; shared situational awareness; situations theory.

Council (NRC) Postdoctoral Fellow at the Air Force Research Laboratory at Wright-Patterson AFB. He holds two PhD degrees, in Particle Physics, and Engineering Management. He teaches decision making in complex situations, and project management, and has published articles in Nature, Nature Communication, IEEE, and other leading journals and conferences.

Mustafa Demir is a senior quantitative research scientist at M2 Research Enterprise in New York and a faculty teaching statistics at Arizona State University (ASU) in Arizona, USA. He received his PhD in Simulation, Modelling, and Applied Cognitive Science, focusing on team coordination dynamics and effectiveness in human-machine teaming from ASU in the Spring of 2017. His specific research interests include human-system integration, operations research, dynamical systems, and quantitative research methods – he has been a member of IEEE since 2015.

# 1  Introduction

Advancements in machine learning algorithms and Artificial intelligence (AI) are making many automated systems more autonomous (Chiou and Lee, 2016). One of the shifts in these technological advancements is the development of adaptive intelligent autonomous systems (IAS) as team members who can interact with other agents (i.e., human or machine) to accomplish a common goal. IAS is "the confluence of Autonomy with Unmanned Systems and AI" in the context of Human-Machine Teaming (HMT) in Command and Control (C2) task environments (Department of the Navy Strategy for IAS, 2021). IAS enables organisations to automate a variety of individual tasks that previously required human-level oversight, abilities, and support. Moreover, accompanying advancements in sensors, onboard processing, and AI programming sophistication are broadening the boundaries of where IAS can strongly operate. For these reasons, we define the system of systems (SoS) level as a combination of both humans and IAS (i.e., HMTs) operating towards a common goal while interacting with an environment. The concept of an SoS composed of humans, machines, and accompanying C2 systems that connect them is commensurate with Maier's (1998) taxonomic distinctions of an SoS and generic collaborative systems. These resulting developments in IAS signify increased effectiveness and greater efficiencies in cognitive processes and performance while reducing the mental and physical workload on human operators (Matthews et al., 2021).

The introduction of IAS as a team member into C2 operations introduces new dimensions of operations with humans operating. However, these new HMT formations are also introducing uncertainty as the SoS level. For instance, IAS and accompanying networked decision support systems could create cognitive overload by introducing an overwhelming number of dynamic environmental factors (e.g., human and machine communications) or presenting anomalous behaviours to humans who are part of this larger SoS. Moreover, a human operator's lack of understanding of IAS's behaviours or outputs may equally produce subsystem uncertainty (Canan et al., 2017). Such algorithmic-induced uncertainty could degrade, slow, or impede decision-making processes, resulting in higher risk to human team members and, in turn, task missions. Current AI research in HMT explores team cognitive processes and performance at the system (i.e., team) level, specifically the SoS level, including interaction (Klien et al., 2004), trust (Bindewald et al., 2018), and decision-making (Kase et al., 2022). However, in the context of human-IAS teams, additional in-depth research, especially in the perspective of team situation awareness (TSA) is needed for engineering these systems for appropriate use.

In general, TSA is defined as "…whereby each team member has a specific set of SA elements about which he or she is concerned, as determined by each member's responsibilities within the team" (Endsley, 1995, pp.38–39). Later, Endsley's TSA perspective was extended by Gorman et al. (2012) by strongly emphasising spatiotemporal aspects of team communication and coordination (i.e., sending information to the right team member, at the right place, at the right time). However, before sending information, the initial decision-making process must also be considered within the TSA concept, especially in the complex domain. To close this gap, the current study aims to understand the dynamics of team-level cognitive processes, including communication, coordination, and decision-making, by exclusively expounding the SoS-level of TSA. Additionally, TSA is more generally related to shared situation awareness (SSA). The research uses SSA and TSA interchangeably in the HMT research (Ososky et al., 2012; Schaefer et al., 2017; Wildman et al., 2014). Therefore, we use SSA throughout the remainder of the paper.

To understand this phenomenon better, this study begins by defining IAS. Next, we examine situational awareness and mental models. We then introduce situations theory to explore how IAS fits within the reality domain perspective (RDP) to better understand SSA among IAS and humans. Finally, by developing an expanded situations theory approach to understanding human-IAS SSA at an SoS level, we outline significant challenges with developing and engineering SSA between humans and IAS. In our final analysis, we demonstrate that IAS lacks requisite variety, the ability to reason counterfactually, and an algedonic feedback loop, and, therefore, cannot generate SSA with human agents. These SoS engineering challenges are discussed and summarised.

## 2 Human and intelligent autonomous systems

IAS technologies pose many new challenges across the spectrum of teaming constructs in sociotechnical systems engineering. Sociotechnical systems engineering takes the principles and methods of sociotechnical systems (e.g., jointly optimising the complex interactions between humans and machines) and applies them to engineering systems (Baxter and Sommerville, 2011). These SoS are composed of numerous subsystems that exhibit an intelligence that is more than the sum of its parts. To understand these systems better, we address several different aspects of IAS intelligence and teaming in this section.

### 2.1 IAS intelligence

The concept of intelligence is subject to a diversity of definitions. Intelligence may include a variety of different aspects within a single definition (Schlinger, 2003). For instance, Clark (2001) points out that human intelligence is hidden in complex and dynamic interactions between technology and the brain (Clark, 2001). The label "intelligent" is often used to describe an interaction behaviour, which creates a circular explanation for the same behaviour (Schlinger, 2003). Stuart Russell (2019) states, "an entity is intelligent to the extent that what it does is likely to achieve what it wants, given what it has perceived" (p.14). Russell's definition provides a performative element based on some goal an entity may hold. However, Russell's definition, like many definitions, enables syllogisms in intelligence discussions. For instance, how does an entity develop "wants" that are "intelligent," or what if it achieves what it wants at the expense of some higher-order virtues? Questions such as these leave most desiring a more comprehensive definition.

A more comprehensive definition by Albus (1991) describes intelligence as "an ability of a system to act appropriately in an uncertain environment, where appropriate action is that which increases the probability of success, and success is the achievement of behavioural sub-goals that supports the system's ultimate goal" (p.474). First, the notion of 'acting appropriately in an uncertain environment' connotes the fulfilment of a system stakeholder's reasonable expectation. However, deviations from a potential range of expected behaviours can engender uncertainty. This definition is also commensurate with INCOSE's description of an open system, which is described as a "system with flows of information, energy, and/or matter between the system and its environment, and which adapts to the exchange" (Sillitto et al., 2019, p.11). Therefore, an important and often overlooked aspect of intelligence in these definitions is the recognition of the environment itself. The environment is concomitant with awareness and understanding, both of which entail change. As a result, including the environment within the definition of intelligence renders it an adaptive notion. In doing so, a distinction between human and machine agent subsystems of the SoS can, in part, be achieved.

While intelligence is usually associated with human behaviour, it is not necessarily limited to only anthropocentric conceptions. Intelligent machines such as computers have augmented human capabilities and have allowed people to perform even more complex work more efficiently. Assemblages of computers into networks, along with associated miniaturisation, have allowed engineers to create advanced systems that can often mimic degrees of human intelligence (Gere, 2009). Producing intelligent machines to replace or relieve humans from the drudgery of dull, dangerous, and dirty occupations is, however, not new (Sharkey, 2008). Yet, defining intelligent machines has proven difficult and is the subject of much debate (Rudas and Fodor, 2008). Intelligent machines often incorporate a range of technologies and methods from the fields of artificial intelligence, cybernetics, operations research, and systems theory (Meystel and Messina, 2000). Intelligent systems consist of a variety of heterogenous components that interact and are interconnected through communicative feedback mechanisms (Merali, 2006). These systems form a complex and integrated whole with clear goals (Choudhury et al., 2019) which are evident in SoS. Yet, these heterogeneous systems can make disambiguating intelligence difficult between the SoS level and the component level. When viewed at a SoS-level, for instance, assemblages of human and machine agents jointly optimised within a framework of a sociotechnical system can manifest emergent intelligence greater than the sum of its parts. However, component-level analysis of individual intelligence contributions may fail to capture the emergent properties of the apparatus of the entire system. Hence, viewing IAS and the human components holistically develops an appreciation for how SSA is based on the interaction these SoS components, and their interaction with the task environment is developed and maintained.

## 2.2 Intelligent autonomous systems

IAS are systems that integrate advanced technologies, often with pattern recognition and learning, that demonstrates intelligence by carrying out tasks in specified environments independent of direct human control and supervision. The U.S. Navy (2021), for instance, defines IAS as "the confluence of autonomy with unmanned systems (UxS) and artificial intelligence (AI)" (p.4). These systems are characterised by their ability to mimic human information processing by alleviating cognitive workloads for systems that exceed human processing capacities (Matthews et al., 2021). Autonomy also implies a level of adaptability that exhibits characteristics of not only self-regulation, but also self-sufficiency (Hoffman et al., 2018). The idea of autonomy here is commensurate with Ireland (2016) who describes autonomy as constituent systems that fulfil the overall purpose or goal of the SoS.

IAS is also an umbrella term that encompasses a variety of technologies. These technologies include self-driving vehicles, the Internet of Things (IoT), the Mars Rover, and drones (Mashkoor et al., 2020). However, groupings of humans and IAS platforms connected via a vast network of supporting technologies form a SoS and may generate a complex situation characterised by different domains of awareness and perspectives. For these reasons, understanding SSA as an ecologically relevant construct amongst humans and IAS technologies is critical for command and control.

## 2.3 Teaming with an IAS

Teams can consist of two or more interdependent agents composed of humans or machines. Although HMT is considered a newer concept, it has been in use with varying levels of interdependency. HMTs can go by several names as well. These range from human-autonomy teams (HATs) (Schaefer et al., 2021) to manned-unmanned teams (MUM-T) (Das et al., 2018). Damacharla et al. (2018) define "HMTs as a combination of cognitive, computer, and data sciences; embedded systems; phenomenology; psychology; robotics; sociology and social psychology; speech-language pathology; and visualisation, aimed at maximising team performance in critical missions where a human and machine are sharing a common set of goals" (p.38637). The comprehensiveness of this definition signifies the complex nature of these advanced systems and the challenges that arise through these interactions. Additionally, what makes the concept of HMT challenging today is the inclusion of intelligent machines and their associated behaviours with higher task interdependency without the capability of eliciting any causal reasoning for its behaviour. More importantly, in the context of SSA between humans and machines, the challenge of negotiating rationality for two (human and machine) as Russell (2019) argues, the intelligent machine in an HMT produces complex behaviours humans cannot rationalise. As a result, humans may not understand or be able to explain a machine's behaviour (Phillips et al., 2011). To this end, an important step to ameliorate human understanding of machine behaviour is having an observable reasoning process based on its external behaviours. Such reasoning has provided leading motives for the development of Explainable AI (XAI) (Jacovi et al., 2021). Yet, there are currently limited theoretical constructs that adequately address teamwork in the context of complex environments with heterogenous teams of humans and intelligent machines (Cooke et al., 2020; Schaefer et al., 2021). The incomprehensibility, concomitant with the absence of an inclusive theory, makes understanding between humans and machines difficult. A great deal of research on situation awareness has attempted to address this, yet, there is still a notable lack of literature on the concept of SSA (Canan and Sousa-Poza, 2019). Situations, such as military operations, may present complex circumstances where IAS and humans may focus on different aspects of information or arrive at different interpretations of the same information. However, SSA requires more than the sharing of information. For example, differences in shared meanings result in contradictory interpretations of battlefield information which are unconducive to SSA (Canan and Sousa-Poza, 2019). Hence, situations may entail different or suboptimal decisions, decision aporia within teams, or poor judgements, all of which may impede timely or appropriate actions at the SoS level.

## 3   Shared situation awareness

Shared situation awareness is an important concept in group decision-making. Because information is what is ultimately shared between agents within an SoS, particular concepts are fundamental to appreciate information's ecological relevance. This below sections introduce several concepts that build on each other to develop a full-orbed understanding the phenomenon and its importance for engineering heterogenous systems of humans and machines.

### 3.1   Shared awareness

Shared awareness is a dynamically adaptive and ecologically relevant cognitive construct that requires the existence of rationality of two or more agents. Alberts et al. (2001) define shared awareness as "a state that exists in the cognitive domain when two or more entities are able to develop a similar awareness of a situation" (p.26). Agents that achieve shared awareness are assumed to have a functional dependency regarding the comprehension of salient aspects within an environment for achieving a particular goal (Canan et al., 2015). Shared awareness is also important for synchronising actions, especially when situations deviate from pre-planned behaviours (Alberts et al., 2001). However, for such sharing to take place, several streams of concepts require convergence to create the concept of shared awareness. Expanding on shared awareness, Nofi (2000) points out that "awareness of a shared situation" implies that a group understands they exist within the same situation; thus, shared awareness of a situation connotes that the group "understands a given situation in the same way" (p.12). Key to awareness also involves a level of transparency of both shared intent and shared awareness (Lyons and Havig, 2014). Yet, understanding a situation in the same way is a complex concept because it implies sharing a single mental model and judgements for how the situation will unfold, given various potential actions. For instance, two agents may share the same social reality, but categorise the phenomenon in two distinct ways as depicted in Figure 1. This discrepancy can therefore result in different actions taken by each agent. However, agents can develop shared awareness in other ways as well.

Alberts et al. (2001) depict four ways two agents can develop shared awareness in Figure 2. However, this depiction of shared awareness is imperfect for several reasons. First, it makes a bid for a three-level ontology. This three-tier ontology makes distinct cuts that remove the nuances of agent interactions with the physical environment. Second, it places the observers outside of the physical realm creating a disconnect between cognitive and physical domains; agents, rather, interact across all three dimensions. Third, it is unclear how two agents, human or otherwise, interact or share a cognitive domain in Alberts's conceptualisation. If one takes the view that cognition can be extended, as Clark and Chalmers (1998) advocate, cognitive artefacts also inhabit the physical world. Alberts et al. model, therefore, provides a confined view of the process of shared awareness that is difficult, if not impossible to achieve. In short, a trifurcated ontology of shared awareness fails to capture how agents interact within situations. These shortfalls demonstrate the need for a more thorough model of awareness to account for developing awareness to better understand if or how a human and IAS may form a truly human-machine team.

**Figure 1** Sharing the same social reality with different understandings (see online version for colours)

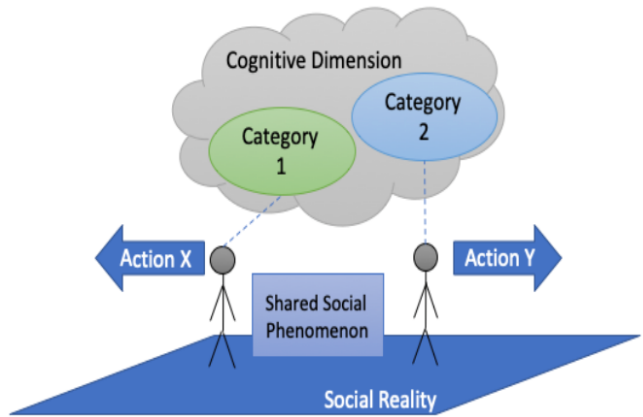

**Figure 2** Achieving shared awareness (see online version for colours)

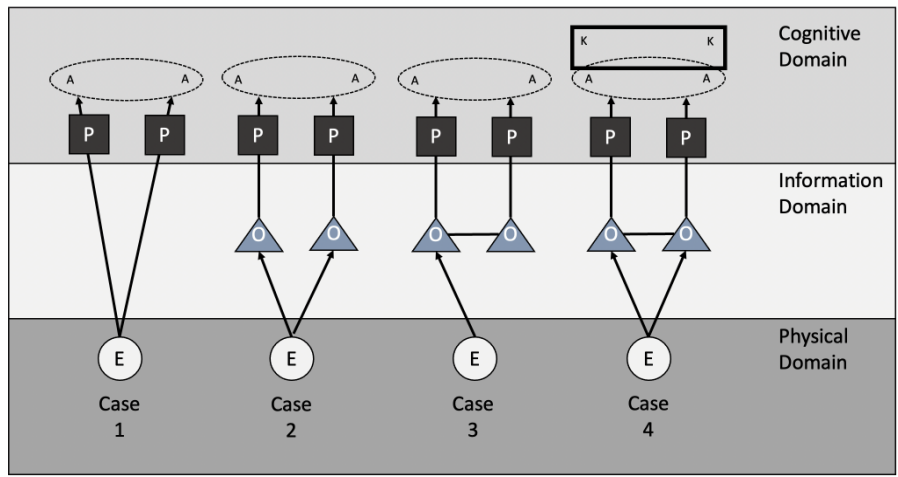

*Source*: Adapted from Alberts et al. (2001)

For these reasons, shared awareness may be partially conceptualised and explicated between human and machine agents through shared mental models.

*3.2 Shared mental models*

Humans often simplify objects and explanations into heuristics via mental models. Mental models represent relationships and sets of assumptions about a system (Ford, 2019). The meanings of mental models are also embedded within a network of other relationships and models (Carley and Palmquist, 1992). In general, mental models provide insight into relevant internal representations that can be activated in situations for comprehension and prediction of the system's future states (Converse et al., 1991). Mental representations contain both declarative and procedural knowledge (Al-Diban, 2012). Procedural knowledge consists of knowledge "know-how" while declarative knowledge is characterised by knowledge of facts and events (ten Berge and van Hezewijk, 1999). Mental models, therefore, combine elements of both procedural and declarative knowledge to instantiate representations of a specific phenomenon within human cognition. Such mental models, while limited in some regards, may also become relatively enduring (Ford, 2019). Additionally, a mental model is an operator's internal representation of a situation or object. A mental model, therefore, allows for the prediction and explanation of outcomes and exerts a major influence on team processes (Matthews et al., 2021). Thus, mental models provide a lens to filter future behaviours to help with prediction.

Mental models afford comprehension of complex situations by extrapolating the state of a system based on the general behaviour of its components. The ability to predict the future behaviour of a system, therefore, requires a comprehensive mental model (Lee and Moray, 1989). Mental models provide a coupling between a system's state and desired goals for allowing a simulation of potential actions (Bryant, 2006; Oury and Ritter, 2021). However, due to information loss through faulty cognitive memory or systems change, a mental model can deviate from reality (Moray, 1996). For example, researchers found that confirmation bias can play a detrimental role in supporting erroneous mental models (Besnard et al., 2004; Kahneman, 2013). While no model is perfect, inaccuracies within models can become more problematic within team settings where agents attempt to develop shared mental models.

Shared mental models are defined by the degree members of a team mutually comprehend a situation (Cannon-Bowers et al., 1993; Klimoski and Mohammed, 1994; Van den Bossche et al., 2011). Shared mental models have been studied in a variety of team contexts, such as cyber security teams (Cotoranu and Chen, 2020), augmented threat detection (Matthews et al., 2021), emergency response management (McNeese et al., 2021), and information systems successes and failures (Carley, 1997). Yet, not all pairings of human and machines are equal. Annett and Stanton (2000) state, a team is a type of group, but not all groups may constitute teams. The key difference hinges on whether team members share a common goal pursued collaboratively (Annett and Stanton, 2000). For instance, the similar concept of shared team cognition is often described in terms of how a team's individual mental models overlap with each other (Cannon-Bowers et al., 1993; Matthews et al., 2021) and, in turn, exhibit high team performance (DeChurch and Mesmer-Magnus, 2010; Johnson-Laird, 1983; Mathieu et al., 2000; Matthews et al., 2021; Rouse et al., 1992). Not surprisingly, Espevik et al. (2006) found that shared mental models added to performance above and beyond operator skill alone. For these reasons, shared mental models have similar roots in other cognitive constructs within the literature.

Perspectives on shared mental models have taken the forms of other cognitive constructs as well. Concepts such as distributed cognition and transactive memory are knowledge focused and act as repositories team members can access to accomplish their goals (Cooke, 2015; Wegner et al., 1985). These concepts, including other boundary objects such as charts, displays, and notes, imply that cognition extends outside of the human mind into the world, which is made accessible to others (Clark and Chalmers, 1998). Similarly, Cooke (2015) states that cognition is often found in checklists, digital artefacts and within the surrounding context of individuals.

If cognition does in fact, extend into shared spaces, then it follows that teams can share cognition through such artefacts. Taking the idea of extended cognition even further, Cooke et al. (2013) proposed a theory of interactive team cognition (ITC). ITC posits that team cognition is made explicit through interactive communications among its members (Cooke, 2015). Likewise, Nofi (2000) states that in sharing individual mental models, "[c]ommunication is the most critical issue in creating shared awareness" and is part of sharing perception of a situation (p.29). ITC, therefore, can be seen as an emergent property of the team when viewed from the SoS level and, therefore, performs two important functions for understanding shared mental models. First, the degree of sharing in mental models can be explicit through communications; second, the degree of sharing between mental models can permit measurement through lexical content analysis, for instance (Carley and Palmquist, 1992; Hutchins et al., 2011). Equally significant, team interactions demonstrate a generative process that gives rise to the importance of how information is shared amongst team members. Yet, it is still unclear how humans and machines could share a mental model when there is still no tractable or computationally shared mental model implemented within a robotic system (Scheutz, 2013; Scheutz et al., 2017). Consequently, this leads to a conceptually impoverished view of SSA of the human-machine teams.

### 3.3 Shared situation awareness

An equally important concept is situation awareness. Tremblay and Banbury (2004), citing Bryant et al. (2004) define "situation awareness (SA) as a cognitive construct that refers to an awareness and understanding of external events in our immediate and near future surroundings" (p.104). The concept, by definition, implies a particular spatial and temporal situation for an agent. Situations can range from simple to complex depending on the interplay between dynamics within a variety of different contexts. Similarly, Endsley (1995) is careful to describe SA as a "state of knowledge" and the process for achieving SA as a "situation assessment" (p.36).

SA is composed of several components. First, Endsley (1995) defines SA as "situation awareness is the perception of the elements in the environment within a volume of time and space, the comprehension of their meaning, and the projection of their status in the near future" (p.36). Endsley's definition consists of three levels of SA:

1 perception

2 comprehension

3 projection.

Level 1 SA (perception) is the process of perceiving essential attributes of the particular task environment (Endsley, 1995). This can be accomplished sensorially, whether it is observed directly or mediated through digital sensors. Level 2 SA (comprehension) involves synthesising elements of Level 1 SA and one's knowledge to create an understanding of a situation (Endsley, 1995). Level 3 SA (projection) entails constituent elements of Level 1 and 2 SA along with knowledge of the spatial-temporal aspects of a situation to forecast a potential future state or its consequences (Endsley, 1995). However, Endsley's model does not specifically address sharing between agents and principally addresses individual SA.

Current models of SA suffer from several shortcomings in conceptualising shared situational awareness. First, Endsley's model is primarily a descriptive theory (Bryant et al., 2004). While descriptive theories are helpful for guiding experimental inquiry, they lack the ability to make predictions and, therefore, provide little explanatory power (Bryant et al., 2004). Thus, Endsley's model lacks explanatory power which hinders relating the SA concept to information processing at the system level. In other words, without an understanding from the process perspective of how individual SA is related to multi-agent informational processing, it is unclear what influence it ultimately plays in accessing decision-making and performance at the SoS level. Second, Endsley's SA model makes it an isolated phenomenon disconnected from more encompassing team processes, such as interdependency and relative information gain. Bryant et al. (2004) argue:

> "Because command and control is not an individual task but rather an intricate set of organisational procedures, we must finally also consider SA as an organisational framework. That is, as well as being the situational representation of the battlespace, SA exists in the organisational structure used to assign information processing roles, decision-making authority, and responsibilities to push information needs down to the sensor level and pull relevant information up to the decision-making units." (p.111)

In other words, SA is a dynamically collective phenomenon inextricably linked to the interaction between components and the task environment. Teams can achieve SSA by beginning with a shared awareness via team interaction (i.e., communication and coordination), mental models of the environment, tasks at hand, and common goals. When teammates share a situation about the current or future state of the system, the degree of similarity between mental models, and knowledge about a mission and tasks, are critical to improving team performance (Mathieu et al., 2000; Schaefer et al., 2021). Similarly, Salas et al. (2005) posit that team adaptability requires members to have a global perspective of the team tasks, understand how changes may alter a team member's role in relation to the team tasks, and be able to recognise changes in other team members, the task, or the environment as they occur. Recognition of team, task, teamwork, and the environment's dynamics implicitly assumes an awareness to glean changes. However, physical separation may inhibit awareness through noise within the environment or communication breakdowns. Team SA (TSA) would likely become more challenging as multiple perspectives increase the dimensions and degrees of dependencies (Dekker and Lützhöft, 2004). To this end, SSA differs from shared awareness of a situation. Shared awareness of a situation, therefore, carries an important nuance, which requires yet further elaboration.

## 4 Expansion of situations theory

Hitherto studies on shared awareness considered shared mental models and SSA. However, these concepts have not provided a comprehensive understanding of what exactly is being shared. To share an understanding *in toto* requires going beyond sharing of information but also includes shared awareness of a situation (Canan and Sousa-Poza, 2019). Moreover, sharing implies mutual communication and coordination for an understanding of particular aspects of reality for the purpose of achieving some mutual goal or end-state (Gorman et al., 2020). Yet, the question remains, how is understanding generated in the first place? To apprehend how agents can share understanding or awareness for behaving intelligently, Situations Theory (ST) is introduced (Sousa-Poza, 2013).

Intelligence requires consideration of the agents and the environment defined by a particular situation. Dekker and Lützhöft (2004) define a situation as "a nested set of constraints that have the potential to shape performance" (p.44). Situations often involve multiple agents and dynamic environments that yield a complex situation. For this reason, ST helps facilitate understanding of complex situations insofar as the perspectives it generates address the encountered problems (Sousa-Poza, 2013). ST defines a situation as "a set of conditions that human beings expand on with the requirement that an individual is or becomes cognizant of the set of the condition(s)" (Canan et al., 2015, p.267). Consequently, ST is a meta-theoretical construct that considers the surrounding conditions of a problem while focusing on understanding the generated agential perspectives for addressing problems within a situation (Canan et al., 2015). Key to understanding ST is the construction of the RDP model.

### 4.1 Reality, domain, and perspective model

The RDP model considers three levels developed through awareness of the self and an other-than-self in a system, e.g., a team (Canan et al., 2015). In ST, the self refers to a self-reflecting agent who can represent the thoughts, perspectives, and intentions of another agent (e.g., theory of mind). The other-than-self refers to other agents that the self seeks to share a situation with (human or machine). However, depending on the order in which one arrives at an awareness of self or other than self, such different perspectives may change how one perceives a particular reality (e.g., order effects). To circumvent this primacy problem, Canan et al. (2015) proposed to view this conundrum as a generative process (GP). A GP resembles a rhizomatic process that has no one starting point. Rhizomatic processes do not follow a linear evolution but is rather an acentric multiplicity that does not develop from a single starting point (de Freitas, 2012). Therefore, we expand on the previously proposed concept of a GP with the additional introduction of liminality.

### 4.2 Generative process

A GP describes a system state in continuous transition. A GP, therefore, lies within a space of liminality, which is between a previous state and a future state of a system. The GP is a subsystem of the SoS because it is related to human cognition as an agent forms awareness of self and other than self (Canan et al., 2015). A GP instantiates a liminal space through an evolving process that results from the onset of a problem. The onset of a problem generates a domain of self-awareness in an agent which is an abstraction from reality. A domain of self-awareness is then sustained and maintained by an agent's cognition as an agent balances the exploration and exploitation of information within the environment. The domain of awareness can thus expand and contract as information is sought, found, compared, and discarded through this system subprocess. As a result, the domain of awareness occupies a continuous space of liminality (e.g., state of continuous transition). Yet, as part of the larger SoS environment, the domain of awareness exhibits dissipative characteristics.

Domains of awareness can be viewed as dissipative system which are systems far from an equilibrium point and yet, maintains a systems' stability (Capra, 1996). Dissipative structures also receive their sources of energy from exogenous sources and can amplify feedback which may cause the system to evolve (Anderson and Meyer, 2016; Capra, 1996). Since agents within a system or team are part of their task environments, they continuously process information both implicitly and explicitly. Agents interact with their environment to maintain a continuous influx of salient information to maintain the structure of a domain of awareness to prevent it from collapsing. However, domains of awareness are not synonymous with the environment, but is rather an abstraction from the environment. An abstraction distance represents the degree of induced complexity (Canan and Sousa-Poza, 2016). If domain of awareness no longer takes in new information to maintain its structure (e.g., maintaining congruency and coherence with the environment), the domain of awareness becomes incoherent, which may result in a steady state system, but one that fails to correctly model reality. In other words, when a domain of awareness no longer resembles reality, the domain of awareness becomes incongruent with the environment. Domains of awareness are, therefore, ephemeral and require the addition of relevant information on a periodic basis to maintain a coherent and congruent understanding of the environment and the problem. Most importantly, domains of awareness also sustain a second-order structure called a perspective.

*4.3 Perspectives*

Perspectives are abstractions within a domain of awareness and represent what is understood of the situation (Canan et al., 2015). Kim et al. (2019) state that perspectives are also viewpoints to which agents may assign different weights to concepts that are the most relevant foci. Perspectives can therefore behave as basins of attraction for information within a domain of awareness as perspectives necessitate information gathering. Additionally, perspectives may also influence the selection of mental models that provide the framework for organising information from the environment. In Figure 3, the congruency of the perspective to reality is judged by an abstraction distance, $A'(d)$. $A''(d)$ denotes what is understood from what is comprehended. The complexity of the situation is designated by $A(d)$ which is a combination of $A'(d)$ and $A''(d)$. As information enters the system, the distances (d) will fluctuate and influence change in the other distances as well. For instance, when an agent conducts exploration, they are attempting to decrease $A'(d)$. Conversely, when an agent conducts exploitation, they are attempting to decrease $A''(d)$. Agents within this system will expend resources and time on both exploration and exploitation to decrease $A(d)$ by decreasing $A'(d)$ and $A''(d)$ to take an action. Once an action is taken, a new state is reached which generates a new domain of awareness. As a result, abstractions and perspectives are reciprocal phenomena which mutually emerge through the process of interacting with the environment. This conceptualisation of relating information exploration to reducing A'(d) and information exploitation to reducing A"(d) is an augmentation to current ST that has not existed before.

**Figure 3** The domain of awareness and perspective with respect to reality

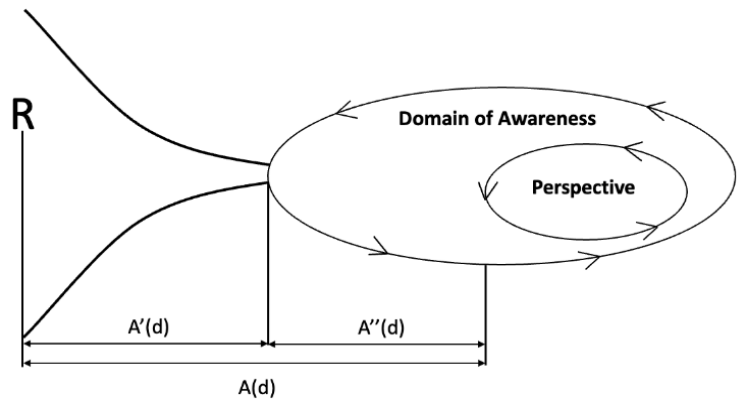

*Source*: Adapted from Canan et al. (2015)

New states within the domain of awareness are reached through actions which could result from conjectural statements, or claims (Canan et al., 2015). These conjectures become the models which are viewed as a kind of hypothesis test. Subsequently, what is understood becomes determined by acting upon these models and the feedback one receives. Feedback then helps determine the usefulness of the models. These models are typically tested through the act of human cognition, namely through scenarios supported by counterfactual thinking.

Counterfactual thinking is fundamental to being human and to understanding in general. Pearl and Mackenzie (2018) state that counterfactuals are building blocks of scientific thought. Counterfactuals are events that have not taken place, but through imagination, a human agent rationalises a future or past that could or would be different (Starr, 2021). In other words, counterfactuals are conceptions of past or future events that may have turned out differently if certain conditions were met, *ceteris paribus*. For example, the nature of scientific hypothesis testing asks "what if" questions which are formulated into testable, falsifiable experiments. Similarly, when humans develop conjectural statements or claims, they are engaging in a rudimentary form of hypothesis formation or development of a provisional counterfactual. To understand this interplay better, we expand ST by adding a domain of counterfactuals, thus introducing the Expanded Situations Theory.

### 4.4 *Expanded situations theory*

Expanded Situations Theory (EST) amends ST by the addition of the counterfactual domain. Counterfactuals are informed by an agent's history (hysteresis), experiences, training, and attitude. More importantly, counterfactuals are used for making plans and assessing outcomes, which are essential for learning and prediction (Narayanan, 2010). As Weick (1985) states, "[t]o go beyond detail is to move to higher levels of abstraction and to invoke alternative realities" (p.61). Therefore, the counterfactual domain is essential for understanding shared awareness for three reasons. First, the counterfactual domain is fundamental to assessing mental models. As a domain of awareness is developed, the generated perspectives will favour certain mental models that may best explain a phenomenon. Succinctly put, a perspective selects a particular model of behaviour that best explains the data. Subsequent actions are then taken by the system agent to confirm or update the mental model. Simulating these mental models based on the data likewise generates counterfactual thinking. This process continues until a sufficient reason for action is taken or the situation changes. Feedback effects from actions also update an agent's mental models for future use. Second, counterfactuals are essential for understanding causal models of the world. Sharing a causal model implies that two agents also share the same counterfactual judgements (Pearl and Mackenzie, 2018). Sharing judgement is, therefore, sufficient for developing consensus and making decisions in a team setting, which are essential for positive team behaviours. Third, the counterfactual domain represents a realm of possible states, with some being more ideal than others. As Dewey (1933) states, "[t]o grasp the meaning of a thing, an event, or a situation is to see it in its relations to other things: to see how it operates or functions, what consequences follow from it, what causes it, what uses it can be put to…" (p.137). Stated differently, counterfactual reasoning creates a space to cognitively simulate a system to help understand the outcomes of decisions. Understanding through counterfactuals thus allows agents to act rationally within their systems. The counterfactual domain thus co-generates the perspective with reality to create, bound, and bind a domain of awareness. The counterfactual domain, therefore, holds the domain of awareness in tension with reality and can drive actional behaviours of information exploration and exploitation. Figure 4 shows the amended RDP model with the counterfactual domain represented by "C". The addition of the counterfactual domain not only bounds the abstraction distance but provides a necessary supplement to assessing differences between human and IAS agents within team settings requiring shared awareness within an EST.

**Figure 4**   Expanded situations theory (EST)

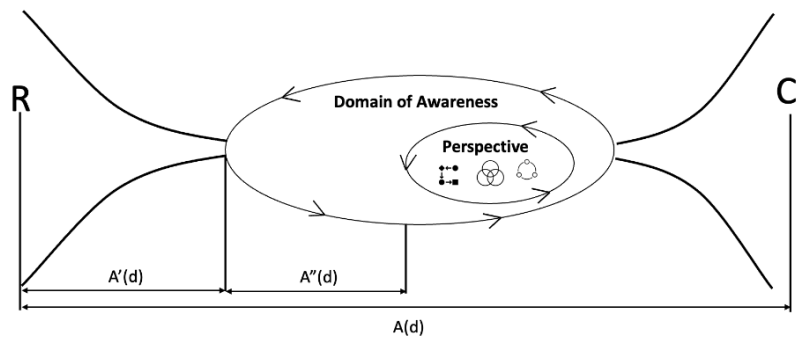

*Source:*   Adapted from Canan et al. (2015)

## 5   Three areas of investigation

Expanded situations theory can provide a meta-theoretical framework for understanding human and IAS interactions in complex task environments, which require proactive interaction between the team members (either human or machine) to attain a shared understanding of the situation (Atkinson et al., 2014). On this more solid footing of EST, three areas that present challenges to achieving shared awareness with IAS:

1 requisite variety in IAS

2 counterfactual reasoning in IAS

3 algedonic feedback loops within the system.

*5.1 Requisite variety in IAS*

Developed by Ashby (1991), the law of requisite variety simply states that a system should exhibit a number of different responses to address inherent complexity within any particular environment. Thus, only a system's variety in the number of possible responses can delimit the number of potential system outcomes (Beer, 1984). In other words, a system must have an equal or greater variety than the number of environmental perturbations to avoid failure or collapse. When addressing shared awareness between human and IAS agents, requisite variety plays a vital role in determining the possible outcomes a human-machine system may exhibit when developing a domain of awareness or shared awareness.

Developing and sustaining a domain of awareness requires an adequate amount of information from the task environment in which the agents are subject to. Information may be gathered through a variety of different sources, such as videos, reports, news feeds, and the like. However, functions of IAS often operate at run time which makes it subject to dynamic inputs from the environment (Johnson and Verdicchio, 2017). While humans can consume a variety of different media in different ways to build a domain of awareness, IAS platforms may be limited. For example, autonomous systems such as IAS platforms will have a limited array of sensors or are equipped with only a small number of different actuators that inherently bounds the range of what it can receive and, and by extension, what it can perform (Johnson and Verdicchio, 2017). These limitations accentuate three issues for achieving shared awareness when it comes to requisite variety. First, human agents will not know the variety of possible inputs an autonomous platform will actually sense and, therefore, not know how the platform may respond, thus making it less rational (Johnson and Verdicchio, 2017). This lack of understanding thus limits what a human agent can possibly understand. Second, if IAS platforms provide inputs to decision support systems such as target identification in military operations, which also populates a common operational picture, the IAS's algorithm filtering may prevent or suppress any number of other important environmental attributes, as portrayed in Figure 5. Limiting to only what IAS platforms can detect, may result in an incomplete picture of reality, which can lead to poor decision-making. Third, intelligent systems are computationally constrained, which determines their range of behaviours (Pothos and Pleskac, 2022). Such computational constraints can potentially limit rational behaviours and inhibit timely decision making. Overall, the results are two incongruent domains of awareness and constrains an agent's (e.g., IAS) requisite variety.

Resolving two different domains of awareness between a human agent and an IAS may result in several other unintended consequences. For instance, human agents can directly limit or constrain an IAS's autonomy by attempting to resolve uncertainty within an environment through redirecting an IAS platform to focus on a particular area.

Through such actions, time is now spent supervising autonomy over exploiting or synthesising information. Moreover, new interventions from human agents could potentially introduce new or unpredictable system behaviours which require human agents to have an unanticipated requisite variety (Behymer and Flach, 2016). Yet, shared awareness still may be unachievable due to the technological mediation of the environment. Weick (1985) points out that for "[systems] to register and absorb information (to listen and think), the sensor must be at least as complex as the information it is receiving and, often, information systems fall short" (p.61). Moreover, Canan et al. (2022) state: no extra information can be obtained to reduce ontic uncertainty; it can only be resolved when the system interacts with the environment. Therefore, if the systems are unable to truly capture the complexity and a human agent, as part of the more extensive system, IAS cannot directly interact with the environment, then shared awareness is not genuinely possible. As a result, IAS has a constrained requisite variety and could introduce the need for additional variety not previously foreseen, which ultimately prevents shared awareness.

**Figure 5**    Constrained domain of awareness

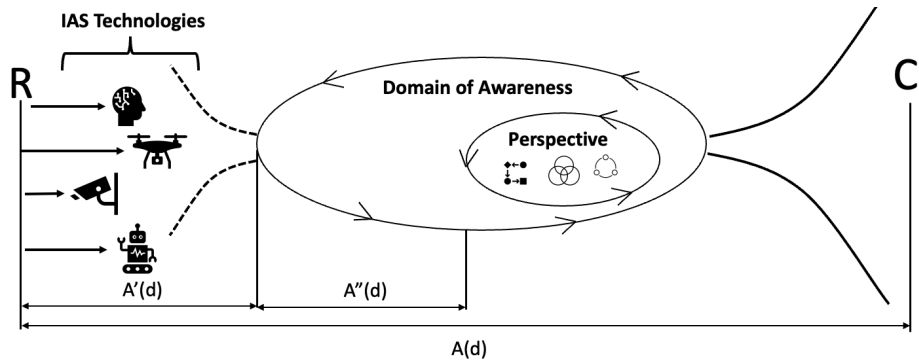

## 5.2  *Counterfactual reasoning in IAS*

Counterfactual reasoning allows agents to contemplate what-if types of questions that help sustain the domain of awareness and the types of perspectives that could be adopted. The very notion of planning or simulating implies a rational intention to proceed from a current state to a desired future state by manipulation of system variables or the environment. As van de Poel (2020) states:

> "The fact that human agents can take an external perspective and can reflect implies that they are able to improvise in unexpected situations and can therefore contribute to the proper functioning of the system and the achievement of certain values which would not have been realised without their intervention." (p.397)

However, it is known that computers today are largely relegated to the "seeing" rung of Pearl and Mackenzie's (2018) ladder of causation. Figure 6 demonstrates how SA maps to Pearl and Mackenzie's (2018) ladder of causation. Without the ability to reason counterfactually, IAS agents' communication and coordination to human agents can suffer. Counterfactual reasoning provides the necessary "why" for prospective actions and actions not taken. Nevertheless, this may be an unreasonable requirement in environments that have many dynamics and levels of understanding, such as military, environmental, and political objectives that must be considered. For this reason, attempting to bridge the shared awareness gap between a human agent and IAS will suffer from several subtle sub-goals as well.

**Figure 6** Levels of situational awareness and ladder of causation

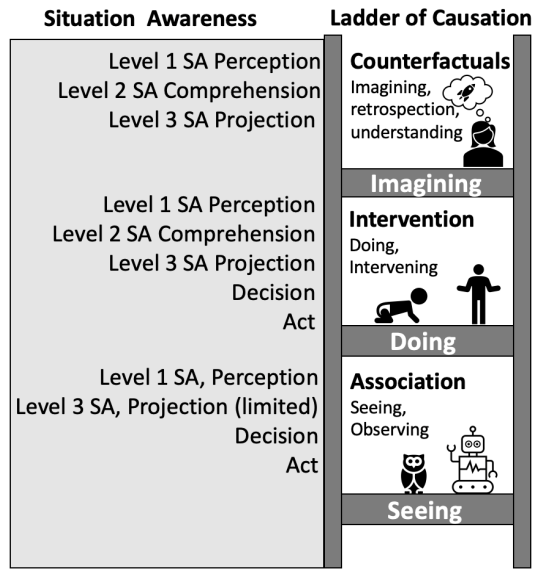

*Source*: Adapted from Pearl and Mackenzie (2018)

First, the autonomy of a machine will likely be a composition of many different types of code bases. For instance, some parts of the system will be deterministically established at compile time, while other components that drive behaviour are established at run-time (Johnson and Verdicchio, 2017). This makes autonomy a system composed of an admixture of deterministic and non-deterministic components that can make understanding behaviours difficult. Moreover, behaviours that entail judgement or intuition fundamentally lack a formal mathematical representation, thus making it extremely difficult to code for within the system (Cummings, 2014). Second, these formal rationalities exhibited by intelligent machines are not only deterministic, but can also suppress a human's substantive rationality needed to make better decisions (Lindebaum et al., 2020). Third, high levels of consistency within these systems have also been shown to be detrimental to performance. For instance, an aerial drone equipped to autonomously take off from the same location unintentionally created divots in the runway overtime until one day, the programmed thrust was not able to move it from this position, which resulted in a flight cancellation and an a detailed investigation into the system (McLean, 2022). In a similar case, autonomous mining trucks had to be programmed to vary their routes over time to avoid creating ever-deepening tracks in the road that made it difficult for other vehicles to operate (Cummings, 2014). These two examples demonstrate that a lack of counterfactual thinking may result in behaviour that becomes too predictable because it lacks the foresight to anticipate suboptimal consequences. While such consistent behaviour may seem beneficial for shared awareness, it is not always the case. Rather, shared awareness implies intentionality and counterfactual reasoning that anticipates how rote actions may result in undesirable system behaviours at higher system levels. Therefore, without the ability to reason counterfactually, it will be improbable for human agents and IAS to develop shared awareness.

### 5.3 *Algedonic feedback loops within the system*

All systems require feedback mechanisms to drive behaviours toward established goals. Feedback is simply information but can manifest in many different forms. Of interest are feedback loops of consequence or those which could particularly result in high uncertainty in mission failure or significant injuries to humans. Principally, algedonic feedback are cybernetic loops that signal pain or pleasure upon a part of the system (Beer, 1972). Pain or pleasure in this instance, may not necessarily be physical but could encompass the mental as well. For instance, a human soldier experiencing mental anguish by coming under the threat of an enemy weapons system, can experience algedonic feedback (e.g., existential anxiety, duress) until the threat is removed. Algedonic feedback can therefore regulate a system by offering a reward or punishment to normalise the system toward its goals. Developing and maintaining shared awareness between human and IAS agents would require both entities share in an algedonic feedback loop. However, IAS cannot possess a true algedonic feedback because it does not experience an existential crisis or risk as a human would, which is an important part of shared awareness.

Humans operating IAS comprise a sociotechnical system in which the components might receive, process, and output information to achieve a common goal(s) of a team task (Galbraith, 1974). Such a system would also contain a number of various feedback mechanisms which help drive acceptable or beneficial behaviours by all members. However, it is well-documented that autonomous systems without a sense of "skin-in-the-game" could result in a number of unanticipated system behaviours that unfavourably affect only human agents in the system (Larsson et al., 2020; Solovyeva and Hynek, 2018). For example, IAS such as autonomous weapons systems (AWS) could remove humans from physical harm, but could also result in a minimisation of human life or turn combat into an "unempathetic automated industrial process" (Solovyeva and Hynek, 2018, p.177). Systems such as IAS that have elements of AI are unlikely to cope with the complex, ever evolving environment while simultaneously trying to support human intentions (Payne, 2018). Lyons et al. (2021) underline that most pilots would not countenance embarking on a dangerous mission with an autonomous wingman because of the fears and uncertainty it raises. Additionally, IAS would not come under punishment for choices it was programmed to make or direct (Larsson et al., 2020). In all cases, the human shares in a feedback loop that IAS cannot be programmed to accommodate, and therefore, cannot truly share awareness.

## 6   Discussion and future research

The development of shared awareness between a human and IAS agents present many challenges. Much of this stems from the fact that autonomous platforms inherently produce *de novo* outcomes (Kallinikos et al., 2013). Novel effects make it difficult to develop true shared awareness due to both the limited rationality, unexplainability, and unpredictability of current machine autonomy, including the concomitant uncertainty the developed within the human agent. Without established shared awareness between an opaque system and a human agent, planning, monitoring, and common goal achievement could suffer or come to cross purposes at points in time if not engineered comprehensively. Consequently, such considerations will likely drive changes in other parts of the system and to higher system levels.

The development of shared awareness includes changes within higher SoS levels. Many current theories of technology and work are too course-grained to capture the nuances of technical change (Orlikowski and Scott, 2008). For instance, next-generation systems engineering must consider how augmentation influences shared awareness among heterogeneous members of a hybrid team (Canan, 2017) and digitally model human-machine interactions (Huang et al., 2020). The development of sociotechnical systems within modern organisations will bring many challenges that are not neutral. Anytime a new technology is introduced in a social context, there are always consequences that require further engineering and analysis (Selbst et al., 2019). We, therefore, must avoid sociotechnical blindness caused by ignoring the human component in attempts at engineering a completely self-sufficient AI; rather, human agents must be at the centre of the design efforts (Johnson and Verdicchio, 2017).

Systems engineering must take a holistic approach that emphasises designing a seamlessly integrated human and technological system (Behymer and Flach, 2016). Automation should not be looked at as comparable or as a replacement for human capabilities, but should be seen as harmonising the capabilities of both (Hoffman et al., 2018; Zahedi and Kambhampati, 2021). Future systems engineering research could develop a mission engineering approach that appropriately orders the myriad perspectives into a more cohesive whole (Sousa-Poza, 2015). As a result, IAS could become a complementary system component rather than a replacement for human agents for developing awareness at the SoS level.

## 7  Summary

This paper takes and discusses the position of developing an expansion to situations theory in the context of human-IAS teams. The further development of a domain of awareness as a generative process provides additional insights and considerations for how shared awareness is formed and sustained over time to achieve complementary goals between IAS and humans for improved performance at the SoS level. Understanding the role of IAS within human-IAS teams still requires a great deal of more research from the perspective of the shared awareness concept (Yan et al., 2021). Such complex assemblages of humans and IAS might suffer from what Cummings calls (2014) "the curse of dimensionality" and not have any "closed-form solutions and will be intractable from a mathematical perspective" (p.68). As argued by Keating and Katina (2012), the field of SoS engineering must garner a greater appreciation for how information technologies are rapidly introducing new complexities. Consequently, the study of IAS is likely to involve a strong systems engineering approach with an emphasis on interdisciplinary perspectives to address these challenges. Systems engineering must develop new insights to further progress in the development of new approaches for engineering a SoS composed of human-machine teams.

The introduction and use of IAS portend manifold engineering, organisational, and cognitive challenges for achieving shared awareness. With the US military as a primary supporter and promoter of human-IAS teams, much of the research will likely center around dynamic C2 task environments (Lyons et al., 2021). In fact, C2 within a military context demonstrates the importance of shared awareness at a team level with IAS. Yet, such dynamic environments may present some of the most critical complex situations with novel challenges for human comprehension of IAS behaviours and subsequent decision-making that may equally apply to the private sector.

As stated in this paper, shared awareness is not a synonym for SSA or information sharing. Rather, shared awareness, while encompassing situation awareness and information sharing, also includes sharing a domain of awareness generated through participating and interacting in a situation with equal comprehension, risks, and goals. Yet, IAS without a requisite variety, the ability to reason counterfactually, or an algedonic feedback loop, shared awareness between humans will likely be an unbridgeable chasm for the near future. Nevertheless, concepts such as digital engineering and mission engineering can potentially mitigate these shortcomings by providing the bridging solutions necessary for organisations to remain competitive while operating safely and effectively.

*Expansion of situations theory for exploring shared awareness* 23